\documentstyle[twoside,fleqn,espcrc2,epsf]{article}

\newcommand{\be}{\begin{equation}}
\newcommand{\ee}{\end{equation}}
\newcommand{\nostrocostrutto}[2]
  {\mathrel{\mathop{\kern 0pt \rlap
  {\raise.2ex\hbox{$#1$}}}
  \lower.9ex\hbox{\kern-.190em $#2$}}}

\newcommand{\AmS}{{\protect\the\textfont2
  A\kern-.1667em\lower.5ex\hbox{M}\kern-.125emS}}

\title{Non--Baryonic Dark Matter}

\author{V. Berezinsky$^1$, A. Bottino$^{2,3}$ and G. Mignola$^{3,4}$ 
\ \\ 
$^1$INFN, Laboratori Nazionali del Gran Sasso, 67010 Assergi (AQ),
Italy \\ 
$^2$Universit\`a di Torino, via P. Giuria 1, I-10125 Torino, Italy \\
$^3$INFN - Sezione di Torino, via P. Giuria 1, I-10125 Torino, Italy \\
$^4$Theoretical Physics Division, CERN, CH--1211 Geneva 23, Switzerland 
\ \\
\vspace{.5truecm}
(presented by V. Berezinsky)       
}

\begin{document}

\begin{abstract}
The best particle candidates for non--baryonic cold dark matter are reviewed, 
namely, neutralino,
axion, axino and Majoron. These particles are considered in the context of
cosmological models with the restrictions given by the observed mass spectrum 
of large scale structures, data on clusters of galaxies, age of the Universe
etc. 

\end{abstract}

\maketitle

\section{\bf Introduction (Cosmological environment)}
Presence of dark matter (DM) in the Universe is reliably established. 
Rotation curves in many galaxies provide evidence for large halos 
filled by nonluminous matter. The galaxy velocity distribution in clusters 
also show the presence of DM in intercluster space. IRAS and POTENT 
demonstrate the presence of DM on the largest scale in the Universe. 

The matter density in the Universe $\rho$ is usually parametrized in 
terms of $\Omega=\rho/\rho_c$, where 
$\rho_c\approx 1.88\cdot10^{-29}h^2~g/cm^3$ is the critical density and 
$h$ is the dimensionless Hubble constant defined as 
$h=H_0/(100 km.s^{-1}.Mpc^{-1})$. Different measurements suggest generally
$0.4\leq h \leq 1$. The recent measurements of extragalactic 
Cepheids in Virgo and Coma clusters narrowed this interval to 
$0.6 \leq h \leq 0.9$. However, one should be cautious about the accuracy of 
this interval due to uncertainties involved in these difficult measuremets.

Dark Matter can be subdivided in baryonic DM, hot DM (HDM) and 
cold DM (CDM).

The density of baryonic matter found from nucleosynthesis
is given \cite{St94} as $\Omega_b h^2=0.025\pm0.005.$

Hot and cold DM are distinguished by velocity of particles at the moment when 
horizon crosses the galactic scale. If particles are relativistic they are 
called HDM particles, if not -- CDM. The natural candidate for 
HDM is the heaviest neutrino, most naturally 
$\tau$-neutrino.
Many new particles were suggested as CDM candidates.

The structure formation in Universe put strong  restrictions to  the properties 
of DM in Universe. Universe with HDM plus baryonic DM  has 
a wrong prediction for the spectrum of fluctuations as compared with 
measurements of COBE, IRAS and CfA. CDM plus baryonic matter can explain the 
spectrum of fluctuations if total density $\Omega_0 \approx 0.3$.
   
There is one more form of energy density in the Universe, namely the 
vacuum energy described by the cosmological constant $\Lambda$. The 
corresponding energy density is given by
$\Omega_{\Lambda}= \Lambda/(3H_0^2)$. Quasar lensing and the COBE results 
restrict the vacuum 
energy density: in terms of $\Omega_{\Lambda}$ it is less than 0.7.

Contribution of galactic halos to the total density is estimated as 
$\Omega \sim 0.03 - 0.1$ and clusters give $\Omega \approx 0.3$.
Inspired mostly by theoretical motivation (horizon problem, flatness problem 
and the beauty of the inflationary scenarios) $\Omega_0=1$ is usually assumed. 
This value is supported by IRAS data and POTENT analysis. No observational data
 significantly contradict this value.

There are several cosmological models based on the four types of DM
described above (baryonic DM, HDM, CDM and vacuum energy). These models 
predict different spectra of fluctuations to be compared with data of  
COBE, IRAS, CfA etc. They also produce  different
effects for cluster-cluster correlations, velocity dispersion etc. The  
simplest and most attractive 
model for a correct description of all these phenomena is the so-called
mixed model or  cold-hot dark matter model (CHDM). This model is
characterized by following parameters:
\begin{eqnarray}
\Omega_{\Lambda}=0, \Omega_0=\Omega_b+\Omega_{CDM}+\Omega_{HDM}=1,\nonumber\\
H_0\approx 50~km s^{-1}Mpc^{-1} (h \approx 0.5), \nonumber\\
\Omega_{CDM}:\Omega_{HDM}:\Omega_b \approx 0.75:0.20:0.05,
\label{eq:chdm}
\end{eqnarray}
where $\Omega_{HDM} \approx 0.2$ is obtained in ref.\cite{Kly} from damped 
$Ly\alpha$ data. Thus in the CHDM model the central value for the CDM density  
is given by 
\begin{equation}
 \Omega_{CDM}h^2 = 0.19  
\label{eq:cdm}
\end{equation}
\noindent
with uncertaities within 0.1.

The best candidate for the HDM particle is $\tau$-neutrino. In the CHDM 
model with $\Omega_{\nu}=0.2$ mass of $\tau$ neutrino is
$m_{\nu_{\tau}} \approx 4.7~eV$. This component will not be discussed further.

The most plausible candidate for the CDM particle is probably the neutralino 
($\chi$):
it is massive, stable (when the neutralino is the lightest supersymmeric 
particle and if R-parity is conserved) and the $\chi\chi$-annihilation 
cross-section results in $\Omega_{\chi} h^2 \sim 0.2$ in  large areas of 
the neutralino parameter space.

In the light of recent measurements of the Hubble constant the CHDM model 
faces the {\em age problem}.
The lower limit on the age of Universe $t_0 > 13$~Gyr (age of globular
clusters) imposes the upper limit on the Hubble constant in the CHDM model 
$H_0 < 50~km s^{-1}Mpc^{-1}$. This value is in slight contradiction with the
recent observations of extragalactic Cepheids,
which can be summarized as $H_0 > 60~km s^{-1} Mpc^{-1}$.
However, it is too early to speak about a serious conflict taking into account
the many uncertainties and the physical possibilities (e.g. the Universe can be 
locally overdense - see the discussion in ref.\cite{Pri95}).

The age problem, if to take it seriously, can be solved with help of 
another successful cosmological model $\Lambda$CDM. This model assumes that 
$\Omega_0=1$ is provided by the vacuum energy (cosmological constant 
$\Lambda$) and CDM. From the limit $\Omega_{\Lambda}<0.7$ and the age of
Universe one obtains  $\Omega_{CDM} \geq 0.3$ and 
$h<0.7$. Thus this model also predicts
$\Omega_{CDM} h^2 \approx 0.15$ with uncertainties 0.1. Finally, we shall 
mention that the CDM with $\Omega_0=\Omega_{CDM}=0.3$ and $h=0.8$, which 
fits the observational data, also gives $\Omega h^2 \approx 0.2$. Therefore 
$\Omega h^2 \approx 0.2$ can be considered as the value common for most 
models.

In this paper we shall analyze several candidates for CDM, best motivated 
from the point of view of elementary particle physics. The motivations are 
briefly described below.\\
{\em Neutralino} is a natural lightest supersymmetric particle 
(LSP) in SUSY. It is stable if R-parity is conserved. $\Omega_{\chi} \sim
\Omega_{CDM}$ is naturally provided by annihilation cross-section in 
large areas of neutralino parameter space.\\
{\em Axion} gives the  best known solution for strong CP-violation.
$\Omega_a \sim \Omega_{CDM}$ for  natural values of parameters.\\
{\em Axino} is a  supersymmetric partner of axion. It can be 
LSP.\\
{\em Majoron} is a  Goldstone particle in spontaneously broken global 
$U(1)_{B-L}$ or $U(1)_L$. $KeV$ mass can be naturally produced by gravitational 
interaction.

Apart from cosmological acceptance of DM particles, there can be  
observational confirmation of their existence. The DM particles can be 
searched for in the direct and indirect experiments. The direct search 
implies the interaction of DM particles occurring inside appropriate  detectors. 
Indirect search is based on detection of the secondary particles produced 
by DM particles in our Galaxy or outside. As examples we can mention 
production of antiprotons and positrons in our Galaxy and high energy gamma 
and neutrino radiation due to annihilation of DM particles or due to their 
decays.
\section{Axion}
The axion is generically a light pseudoscalar particle which gives natural and  
beautiful 
solution to the CP violation in the strong interaction \cite{PQ77}
(for a review and references see\cite{KT90}). Spontaneous breaking of the 
PQ-symmetry due to VEV of the scalar field  $<\phi>= f_{PQ}$ results in the 
production of massless Goldstone boson. Though $f_{PQ}$ is a free 
parameter, in practical applications it is assumed to be large,
$f_{PQ} \sim 10^{10} - 10^{12}~GeV$ and therefore the PQ-phase transition 
occurs in very early Universe. At low temperature 
$T \sim \Lambda_{QCD} \sim 0.1~GeV$ the chiral anomaly 
of QCD induces the mass of the Goldstone boson
$m_a \sim \Lambda_{QCD}^2/f_{PQ}$ . This massive Goldstone particle is the 
{\em axion}. The interaction of axion is basically determined by the Yukawa 
interactions of field(s) $\phi$ with fermions. Triangular anomaly, which 
provides the axion mass, results in the coupling of the axion with two photons.
Thus, the basic  for cosmology and astrophysics axion interactions are those 
with nucleons, electrons and photons. 

Numerically, axion mass is given by
\be
m_a=1.9\cdot 10^{-3}(N/3)(10^{10}~GeV/f_{PQ})~eV,
\label{eq:am}
\ee
where $N$ is a color anomaly (number of quark doublets).

All coupling constants of the axion are inversely proportional to $f_{PQ}$
and thus are determined by the axion mass. Therefore, the upper limits on emission 
of axions by stars result in upper limits for the axion mass. Of course, 
axion fluxes cannot be detected directly, but they produce  additional 
cooling which is limited by some observations (e.g.,age of a star, duration of 
neutrino pulse in case of SN etc). In Table 1 we cite  the upper limits 
on axion mass from ref.\cite{KT90}, compared with revised limits, given 
recently by Raffelt \cite{Raf95}.
\begin{table}[hbt]
\caption{Astrophysical upper limits on axion mass}
\center{\begin{tabular}{|c|c|c|} \hline
                    &1990 \cite{KT90}  & 1995 \cite{Raf95}\\
\hline
\hline
sun                  &$1~eV$                & $1~eV$\\ \hline

red giants           &$1\cdot 10^{-2}~eV$   & \begin{tabular}{l}
                                                very \\
                                               uncertain
                                            \end{tabular} \\ \hline
\begin{tabular}{l}
hor.--branch \\
stars 
\end{tabular}  & \begin{tabular}{l}
                  not \\
                  considered
                  \end{tabular} 
& $0.4~eV$\\ \hline
SN 1987A             &$1\cdot 10^{-3}~eV$   & $1\cdot 10^{-2}~eV$\\
\hline
\end{tabular}}
\end{table}

As one can see from the Table the strong upper limit, given in 1990 from
red giants, is replaced by the weaker limit due to the horizontal-branch 
stars. The upper limit from SN 1987A was reconsidered taking into account 
the nucleon spin fluctuation in $N+N \to N+N+a$ axion emission.

There are three known mechanisms of {\em cosmological production} of 
axions.  They are (i)thermal production, (i) misalignment production and 
(iii) radiation from axionic strings.

The relic density of thermally produced axions is about the same as 
for light neutrinos and thus for 
the mass of axion $m_a \sim 10^{-2}~eV$ this component is not important 
as DM.

The {\em misalignment production} is clearly explained in ref.\cite{KT90}.

At very low temperature $T\ll\Lambda_{QCD}$ the massive axion provides 
the minimum of the potential at value $\theta=0$,which corresponds to 
conservation of CP. At very high temperatures $T\gg \Lambda_{QCD}$ the axion 
is massless and the potential does not depend on $\theta$.  At these 
temperatures 
there is no reason for 
$\theta$ to be zero: its values are different in various casually 
disconnected regions of the Universe. When $T\to\Lambda_{QCD}$ the system 
tends to go to potential minimum (at $\theta=0$) and as a result oscillates 
around this position. The energy of these coherent oscillations is
the axion energy density in the Universe. From cosmological point of view 
axions in this regime are equivalent to CDM. The energy density of this 
component is approximately \cite{KT90,Bat94}

\be
\Omega_ah^2 \approx 2\cdot (m_a/10^{-5}~eV)^{-1.18}.
\ee
Uncertainties of the calculations can be estimated as $10^{\pm 0.5}$.

Axions can be also produced by radiation of {\em axionic strings}
\cite{KT90},\cite{BaSh94}. Axionic string is a one-dimension vacuum defect 
$<\phi_{PQ}>=0$, i.e. a line of old vacuum embedded into the new one. 
The string network includes the long strings and closed loops which 
radiate axions due to oscillation. There were many uncertainties in the 
axion radiation by axionic strings (see ref.\cite{KT90} for a review). 
Recently more detailed and accurate calculations were performed by 
Battye and Schellard \cite{BaSh94}. They obtained for the density of axions
\be
\Omega_ah^2 \approx A(m_a/10^{-5}~eV)^{-1.18}
\ee
with $A$ limited between 2.7 and 15.2 and 
with uncertainties of the order $10^{\pm 0.6}$. The overproduction condition 
$\Omega_ah^2 >1$ imposes lower limit on axion mass 
$m_a> 2.3\cdot10^{-5}~eV$.

\begin{figure*}[t]
\epsfxsize=12truecm
\centerline{\epsffile{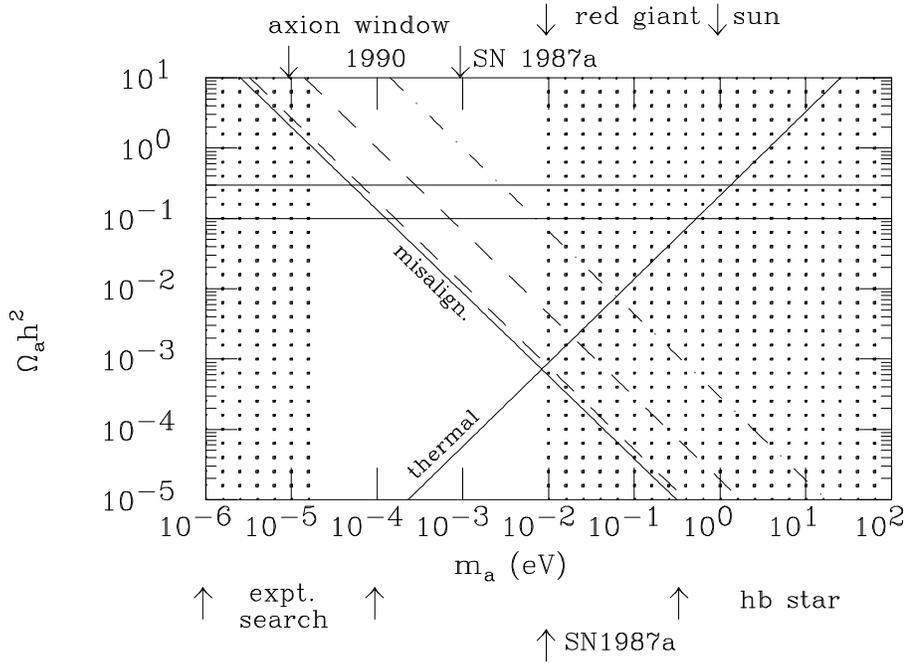}}
\caption{Axion window 1995.
The curves "therm." and "misalign." describe the thermal and misalignement
production of axions, respectively. The dash-dotted curve corresponds to the
calculations by Davis \protect \cite{Dav86} for string production. 
The recent refined calculations \protect \cite{BaSh94} are shown by two dashed 
lines for two extreme cases, respectively. The other explanations are given 
in the text.}
\end{figure*}
Fig.1 shows the density of axions $\Omega_a h^2$ as a function of the 
axion mass $m_a$. The upper limits  on axion mass from Table I are shown 
above the upper absciss (limits of 1990) and below lower absciss (limits 
of 1995). The overproduction region $\Omega_ah^2>1$ and the regions excluded 
by astrophysical observations \cite{Raf95} are shown as the dotted areas.

The axion window of 1995 (shown as undotted region) 
became wider and moved to the right as compared with window 1990.
The  horizontal strip shows $\Omega_{CDM}=0.2 \pm 0.1$ as it was 
discussed in Introduction. One can see from Fig.1 that string and 
misaligment mechanisms provide the axion density as required by 
cosmological CDM model, if axion mass is limited between 
$7\cdot10^{-5}~eV$ and $7\cdot 10^{-4}~eV$. However, in the light of 
 uncertainties, mostly in the calculations of axion production,
one can expect that this "best calculated" window is between
$3\cdot 10^{-5}$ and $10^{-3}$~eV. This region is partly 
overlapped with a possible direct search for the axion in nearest-future 
experiments (see Fig.1 and refs.\cite{axdet}).

\section{Axino}
In supersymmetric theory the PQ-solution for strong CP-violation should be
generalized. Within this theory the PQ symmetry breaking results in the 
production of the 
Goldstone chiral supermultiplet which contains two scalar fields and their 
fermionic partner -- axino ($\tilde{a}$). The scalar fields enter the 
supermultiplet 
in the combination $(f_{PQ}+s)\exp(a/f_{PQ})$, where $s$ is a scalar field,
saxino, which describes the oscillations of the initial field $\phi$ around 
its VEV value $<\phi>=f_{PQ}$, and $a$ is the axion field. This phase
transition in the Universe occurs at temperature $T \sim f_{PQ}$. 
As we saw in the previous section the axion is massless at this temperature 
and since supersymmetry is not broken yet, the axino and saxino are massless,
too. The axion acquires the mass in the usual way due to chiral anomaly at 
$T\sim \Lambda_{QCD}$, while saxino and axino obtain the masses due to 
global supersymmetry breaking. 

The saxino is not of great interest for 
cosmology: it is heavy and it decays fast (mostly into two gluons).
The axino can be the lightest supersymmetric particle and thus another
candidate for DM.

How heavy the axino can be?
The mass of axino has a very model dependent value.
In the phenomenological approach, using the global supersymmetry breaking 
parameter $M_{SUSY}$ one typically obtains (e.g. \cite{TW82},\cite{Nie86}) 
\be
m_{\tilde{a}} \sim M_{SUSY}^2/f_{PQ}
\label{eq:ma}
\ee
For example, if global SUSY breaking occurs due to VEV of auxiliary field of
the goldstino supermultiplet $<F>=F_g$, then the axino mass appears due to 
interaction term $(g/f_{PQ})\tilde{a}\tilde{a}F$ (F has a dimension $M^2$),
and using $<F>=F_g=M_{SUSY}^2$ one arrives at the value (\ref{eq:ma}).

The situation is different in supergravity. In ref.\cite{ChL95} the general
analysis of the axino mass is given in the framework of local supersymmetry.
It was found that generically the mass of axino in these theories is 
$m_{\tilde{a}} \sim m_{3/2} \sim 100~GeV$. Even in case when  axino 
mass is small at tree level, the radiative corrections raise this mass 
to the value $\sim m_{3/2}$. This result holds for the most general form of 
superpotential.The global SUSY result,
$m_{\tilde{a}} \sim m_{3/2}^2/f_{PQ}$, can be reproduced in the local SUSY
only if one of the superpotential coupling constants is very small,
$\lambda <10^{-4}$, which implies fine-tuning. Thus, the axino is too heavy 
 to be a CDM particle.

The only exceptional case was found by 
Goto and Yamaguchi \cite{GY92}. They demonstrated that in case of no-scale  
superpotential the axino mass vanishes and the radiative corrections in 
some specific models can result in the axino mass $10 - 100~keV$, cosmologically
interesting. This beautiful case gives essentially the main foundation for 
axino as CDM particle.

The cosmological production of axinos can occur through thermal production
\cite{RTW91}
or due to  decays of the 
neutralinos \cite{BGM89},\cite{RTW91}. The axion chiral supermultiplet 
contains two particles which can be CDM particles, namely axion and 
axino. In this section we are interested in the case when axino gives the 
dominant contribution. In particular this can take place in the range  
$2\cdot 10^9~GeV <f_{PQ}< 2.7\cdot 10^{10}~GeV$ where axions are cosmologically 
unimportant.

Since axino interacts with  matter very weakly, the decoupling temperature 
for the thermal production is very high \cite{RTW91}:
\be 
T_d \approx 10^9~GeV(f_{PQ}/10^{11}~GeV).
\ee
Therefore, axinos are produced thermally at the reheating phase after
inflation. The relic concentration of axinos can be easily evaluated 
for the reheating temperature $T_R$ as
\be
\Omega_{\tilde{a}}h^2 \approx 0.6 \frac{m_{\tilde{a}}}{100~keV}
(\frac{3\cdot 10^{10}~GeV}{f_{PQ}})^2\frac{T_R}{10^9~GeV}
\label{eq:omega}
\ee
Reheating temperature $T_R \leq 10^9~GeV$ gives no problem with the gravitino 
production. The relic density (\ref{eq:omega}) provides 
$\Omega_{CDM}h^2 \sim 0.2$ for a reasonable set of parameters $m_{\tilde{a}},
f_{PQ}$ and $T_R$. One can easily incorporate in these calculations the 
additional entropy production if it occurs at EW scale \cite{BGM94}.

If the axino is LSP and the neutralino is the second lightest supersymmetric 
particle, the axinos can also be produced by neutralino 
decays \cite{BGM89},\cite{RTW91},\cite{BGM94}. According to estimates 
of ref.\cite{BGM94} the axinos are produced due to 
$\chi \to \tilde{a}+\gamma$ decays at the epoch with red-shift 
$z_{dec}\sim 10^8$. Axinos are produced in these decays as ultrarelativistic 
particles and the free-streeming prevents the growth  of fluctuations on the 
horizon scale and less. At red-shift $z_{nr} \sim 10^4$ axinos become
non-relativistic due to adiabatic expansion (red shift). From this moment on 
the axinos behave as the usual CDM and the fluctuations on the scales 
$\lambda \geq (1+z_{nr})ct_{nr}$ (which correspond to a mass larger than 
 $ 10^{15} M_{\odot}$) grow as in the case of standard CDM.
For smaller scales the fluctuations, as was explained above, grow less
than in CDM model. Therefore, as was observed in ref.\cite{BGM94}, the 
axinos produced by neutralino decay behave like HDM. It means that
axinos
can provide generically both components, CDM and HDM, needed for description of 
observed spectrum of fluctuations.

Unfortunatelly stable axino is unobservable. In case of very weak R-parity 
violation, decay of axinos can produce a diffuse X-ray radiation, with 
practically no signature of the axino.

\section{Majoron}
The Majoron is a Goldstone particle associated with spontaneously broken global
$U(1)_{B-L}$ or $U(1)_L$ symmetry. The symmetry breaking occurs due to VEV of 
scalar field, $<\sigma>=v_s$, and $\sigma$ splits into two fields,
$\rho$ and $J$:
\be
\sigma \to (v_s+\rho)\exp(iJ/v_s).
\ee
The field $J$ is the Majoron. A mass of $\sim$ $keV$ can be obtained 
due to gravitational 
interaction\cite{Akh92},\cite{sec92}. The $keV$ Majoron has the great 
cosmological interest since the Jeans mass associated with this particle,
$m_{Jeans} \sim m_{Pl}^3/m_J^2$,
gives the galactic scale $M \sim 10^{12}M_{\odot}$. In all other respects the 
$keV$ Majoron plays the role of a CDM particle. It is assumed usually that the 
Majoron interacts directly only with some very heavy particles (e.g. with 
the right-handed neutrino $\nu_R$). It results in very weak interaction of the 
Majoron
with the ordinary particles (leptons, quarks etc) and thus makes the Majoron 
"invisible" in the accelerator experiments.

The cosmological production of the Majoron occurs through thermal 
production\cite{sec92,CKO93} and  radiation by 
the strings \cite{sec92} as in case of the
axion. However, under imposed observational constraints, the Majoron in 
models\cite{Akh92},\cite{sec92} has to be in general unstable with lifetime 
 much shorter than the age of Universe. A successful model 
was developed in ref.\cite{BV93}, where the Majoron was assumed to interact 
with the ordinary particles through new heavy particles. For the 
cosmological production it was considered the phase transition associated 
with 
the global $U(1)_{B-L}$ symmetry breaking, when the Majorons were produced both
directly and through $\rho \to J+J$ decays.

Another interesting possibility was considered recently in ref.\cite{DPV95}.
In this model the Majoron is rather strongly coupled with $\nu_{\mu}$ and 
$\nu_{\tau}$ neutrinos ($J\nu_{\mu}\nu_{\tau}$ coupling). The Majorons are 
produced through $\nu_{\tau} \to J+\nu_{\mu}$ decays. The strong interactions 
between Majorons reduces the relic abundance of the Majorons to the 
cosmologically required value.

The Majoron signature in all these models is given by 
$J \to \gamma+\gamma$ decays,
which result in the production of $keV$ X-ray line in the X-ray background 
radiation.

\section{Neutralino}
The neutralino is a superposition of four spin 1/2 neutral fields: the wino 
$\tilde{W}_3$, bino $\tilde{B}$ and two Higgsinos $\tilde{H}_1$ and 
$\tilde{H}_2$:
\be
\chi = C_1\tilde{W}_3+C_2\tilde{B}+C_3\tilde{H}_1+C_4\tilde{H}_2
\label{eq:chi}
\ee
The neutralino is a Majorana particle. With a unitary relation between the
coefficients $C_i$ the parameter space of neutralino states is described by
three independent parameters, e.g. mass of wino $M_2$, mixing parameter of two 
Higgsinos $\mu$, and the ratio of two vacuum expectation values 
$\tan\beta=v_2/v_1$.

In literature one can find two extreme approaches 
describing the neutralino as a DM particle.

(i){\em Phenomenological approach}.
The allowed neutralino parameter space is restricted by the LEP and CDF data.
In particular these data put a lower limit to the neutralino mass,
$m_{\chi}> 20$~GeV. In this approach only the usual GUT relation between  
 gaugino masses, $M_1:M_2:M_3=\alpha_1:\alpha_2:\alpha_3$, 
is used  as an additional assumption, where $\alpha_i$
are the gauge coupling constants. All other SUSY masses which are needed for the 
calculations are treated
as free parameters, limited from below by accelerator data. 

One can find the relevant 
calculations within this approach in refs.\cite{Bott94,Bott95}
and in the review\cite{JKG} (see  also the references therein). There are 
large areas
in neutralino parameter space where the neutralino relic density satisfies the 
relation (\ref{eq:cdm}). This is especially true for heavy neutralinos with
$m_{\chi}>100 - 1000$~GeV, ref.\cite{Tur}. In these areas there are  good
prospects for {\em indirect} detection of neutralinos, due to high energy 
neutrino radiation from Earth and Sun (see \cite{Ka91,Bott95a} and
references therein) as well as due to production of antiprotons and positrons 
in our  Galaxy. The {\em direct} detection of neutralinos is possible too, 
though in more restricted parameter space areas of light neutralinos 
(see review \cite{JKG}).

This model-independent approach is very interesting as an extreme 
case: in the absence of an experimentally confirmed SUSY model it gives
the results obtained
within most general framework of supersymmetric theory.\\
(ii) {\em Strongly constrained models}.
This approach is based on  the remarkable observation that in the minimal 
SUSY SU(5) model with fixed particle content, the three running coupling
constants meet at one point corresponding to the GUT mass $M_{GUT}$. 
Because of the fixed particle content of the model, its predictions are 
rigid and they strongly restrict the neutralino parameter space.
This is especially true for the limits due to proton decay 
$p \to K^{+}\nu$. As a result very little space is left for neutralino as DM
particle. Normally neutralinos overclose the Universe ($\Omega_{\chi}>1$). 
The relic density decreases to the allowed values in very restricted  areas 
where  $\chi\chi$-annihilation is accidentally 
large (e.g.due to the $Z^0$ exchange term - see 
ref.\cite{Nan92}. Thus, this approach looks rather pessimistic for 
neutralino as DM particle.

In several recent works \cite{Nan93}-\cite{RS95} less restricted SUSY
models were considered. In particular the limits due to  proton decay were 
lifted. A GUT model was not specified or less restrictive $SO(10)$ model was  
used \cite{RS95}.
Although, the neutralino can be heavy in these models, the prospects for
indirect detection, including the detection of high energy neutrinos from 
the Sun and Earth, are rather pessimistic \cite{Diel}. The {\em direct} 
detection
is possible in many cases \cite{NA93,AN95,Diel}. 

(iii){\em Relaxed restrictions}. 
In section 5.3, following refs.\cite{Pok95},\cite{Ber95}, we shall analyze the  
restrictions to 
neutralino as DM particle, imposed by {\em basic} properties of SUSY theory.
As in many previous works,  a fundamental element of analysis is 
the radiatively induced EW symmetry breaking (EWSB)\cite{IR82}. However, some
mass unification conditions at the GUT scale are relaxed
(see ref.\cite{Pok95}). The powerful restriction from the no-fine-tuning 
condition is added.
\subsection{\bf SUSY theoretical framework}
The basic element which should be used in the analysis is a  
supersymmetry breaking and induced by it (through radiative corrections)
electroweak symmetry breaking \cite{IR82}. We shall refer to this restriction 
as to the EWSB restriction.

One starts with unbroken supersymmetric model described by some superpotential.
It is assumed that local supersymmetry is broken by supergravity in the hidden
sector, which communicates with the visible sector only gravitationally. This 
symmetry breaking penetrates into the visible sector in the form of global
supersymmetry breaking. More specifically it is assumed that the symmetry
breaking terms in the visible sector are the {\em soft breaking terms} given at 
the GUT scale $Q^2 \sim M_{GUT}^2$ by the following expression:
\begin{eqnarray}
L_{sb}&=&m_0^2\sum_{a}\vert\phi_a\vert^2+m_{1/2}\sum_{a}\lambda_a\lambda_a +
\nonumber \\ 
& & \mbox{}+ Am_0f_Y+Bm_0\mu H_1 H_2
\label{eq:sbt}
\end{eqnarray}
where $\phi_{a}$ are scalar fields of the model (sfermions and two Higgses $H_1$
and $H_2$), $\lambda_a$ are gaugino fields, $f_Y$ are trilinear Yukawa 
couplings of fermions and Higgses and the last term is an additional 
(relative to 
the superpotential term $\mu H_1 H_2$) soft breaking mixing of two Higgses. 
Here and everywhere below we specify the scale at which an expression and
parameters are defined.

The soft breaking terms (\ref{eq:sbt}) are described by 5 free parameters :
$m_0, m_{1/2}, A, B$ and $\mu$. This implies the strong assumption 
that all scalars $\phi_a$ and all gauginos $\lambda_a$  
 at the GUT scale have the common masses $m_0$ and $m_{1/2}$,
respectively. This assumption  can be relaxed, as we shall
discuss later. 

The soft breaking terms (\ref{eq:sbt}) together with supersymmetric mixing give
the following potential defined at the EW scale at the tree level:
\begin{eqnarray}
V&=&m_1^2|H_1|^2+m_2^2|H_1|^2-\nonumber\\
& &\mbox{}-B\mu m_0(H_1H_2+hc)+ \nonumber \\ 
& &\mbox{}+\frac{g_1^2+g_2^2}{8}(|H_1|^2-|H_2|^2)^2,
\label{eq:pot}
\end{eqnarray}
The mass parameters $m_1$ and $m_2$ at the GUT scale are equal to
\be
m_1^2(GUT)=m_2^2(GUT)=\mu^2+m_0^2,
\label{eq:mass}
\ee 
with $\mu$ defined at the GUT scale,too.
The term $\mu^2$ in Eqs.(\ref{eq:pot}),(\ref{eq:mass}) appears due to the 
mixing term $\mu H_1 H_2$ in the superpotential of unbroken SUSY.

The radiative EWSB occurs due to evolution of $m_{H_2}$, the mass of $H_2$, 
which is connected with the upper components of the fermion 
doublets and  
in particular with t-quark. Because of the large mass of t-quark and 
consequently
the large Yukawa coupling $Y_{ttH_2}$, $m_{H_2}^2$ evolves from $m_0^2$ at the 
GUT scale to the negative value $m_{H_2}^2<0$ at the EW scale. At this value 
the potential (\ref{eq:pot}) acquires its minimum and the system undergoes the 
EW phase transition coming to the minimum of the potential. 
At EW scale in the tree approximation the conditions of 
the potential minimum (vanishing of the derivatives) give: 
\begin{eqnarray}
& &\mu^2=\frac{m_{H_1}^2-m_{H_2}^2\tan^2\beta}{\tan^2\beta-1}-
\frac{M_Z^2}{2} \nonumber\\
& &\sin 2\beta = \frac{-2B\mu}{m_{H_1}^2+m_{H_2}^2+2 \mu^2}
\label{eq:ewsb}
\end{eqnarray}

With these equations we obtain one connection between five free 
parameters describing the soft-breaking terms (\ref{eq:sbt}). Thus the number 
of independent parameters is reduced to four, e.g.$m_0, m_{1/2}, A, \mu$
 (or $\tan\beta$).

Using the renormalization group equations (RGE) one can follow the evolution
of the scalar particles (Higgses and sfermions) and spin 1/2 particles 
(gauginos) from the masses $m_0$ and $m_{1/2}$ at the GUT scale to the masses 
at the EW scale. Analogously, the evolution of the coupling 
constants can be calculated. In particlular the masses of Higgses at EW 
scale are given by
\begin{eqnarray}
m_{H_i}^2&=&a_im_0^2+b_im_{1/2}^2+ \nonumber \\ 
& & \mbox{}+ c_iA^2m_0^2+d_iAm_0m_{1/2},
\end{eqnarray}
where $a_i, b_i, c_i $ and $d_i$ are numerical coefficients, which depend on
$\tan\beta$. 

Equivalently, using Eqs.(\ref{eq:ewsb}) one finds
\begin{eqnarray}
M_Z^2&=&J_1m_{1/2}^2+J_2m_0^2+J_3A^2m_0^2+ \nonumber \\
& & \mbox{} + J_4Am_0m_{1/2}-\mu^2,  
\label{eq:nft}
\end{eqnarray}
where $J_i$ are also numerical coefficients.

Eq.(\ref{eq:nft}) allows to impose the no-fine-tuning condition in the 
neutralino parameter space. 
Indeed, one can keep large values of masses in the rhs 
of Eq(\ref{eq:nft}) only by the price of accidental compensation between 
the different terms. It is unnatural to expect an accidental compensation
to a value less than $1\%$ from the initial  values. This 
is the no-fine-tuning condition.\\
\noindent
Naturally this condition is just the same as the one due to the 
radiative corrections to the Higgs mass.
\subsection{\bf Restrictions: the price list}

Within the theoretical framework outlined above one can choose the 
restrictions from the following price list:
\begin{itemize}
\item
Soft breaking terms (\ref{eq:sbt}) and EWSB conditions (\ref{eq:ewsb}),
\item
No-fine-tuning condition,
\item
Particle phenomenology (constraints from accelerator experiments and 
the condition that the neutralino is the LSP),
\item
Restrictions due to $b \to s\gamma$ decay,
\item
Meeting of coupling constants at $M_{GUT}$,
\item
$b-\tau$ and $b-\tau-t$ unification,
\item
Restrictions due to $p \to K\nu$ decay.
\end{itemize}
Using some (or all) restrictions listed above one can 
start  calculations for the neutralino as DM particle. The regions where 
the neutralinos are  overproduced ($\Omega_{\chi}h^2>1$) must be excluded from 
consideration and the allowed region should be determined according to the 
chosen cosmological model (e.g. $\Omega_{\chi}h^2=0.2\pm 0.1$ for the CHDM 
model). For the allowed regions the signal for direct and indirect detection
can be calculated.

\subsection{\bf SUSY models with  basic restrictions}
Accepting all restrictions listed above one arrives at a rigid SUSY model,
 with the neutralino parameter space being too strongly constrained. 
In ref.\cite{Ber95}
the SUSY models with  basic restrictions were considered. These restrictions 
are as  follows:\\
\noindent
(i)Radiative EWSB, (ii) No fine-tuning stronger than $1\%$, (iii) RGE and 
particle phenomenology (accelerator limits on the calculated masses and 
the condition that neutralino is LSP), (iv) Limits from $b \to s\gamma$ decay
taken with the uncertainties in the calculations of the decay rate 
and (v) $0.01 < \Omega_{\chi}h^2<1$ as the allowed relic density for 
neutralinos. Rather strong restrictions are imposed by the condition (ii);
in particular it limits the mass of neutralino as $m_{\chi}<200$~GeV.

At the same time some restrictions are lifted as being too model-dependent:
(i) No restrictions are imposed due to $p \to K\nu$ decay, (ii) Unification 
of coupling 
constants at the GUT point is allowed to be not  exact (it is 
assumed that new very heavy particles can restore the unification),
(iii) unification in the soft breaking terms (\ref{eq:sbt}) is relaxed.
Following ref.\cite{Pok95} it is assumed that masses of Higgses at the GUT 
scale can deviate from the universal  value $m_0$ as
\be
m_{H_i}^2(GUT)=m_0^2(1+\delta_i)\qquad(i=1,2).
\ee
\begin{figure}[htb]
\epsfxsize=6truecm
\centerline{\epsffile{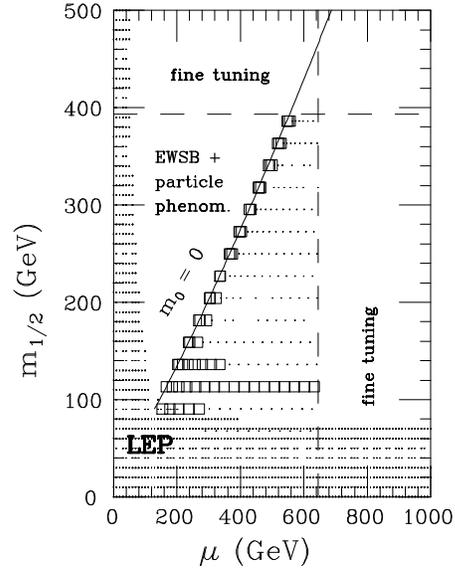}}
\caption{
The neutralino parameter space for the mass--unification case
$\delta_1=\delta_2=0$ and $\tan\beta=8$.} 
\end{figure}

\begin{figure}[htb]
\epsfxsize=6truecm
\centerline{\epsffile{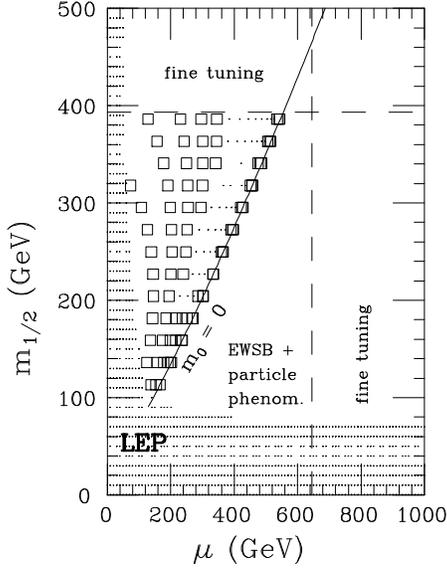}}
\caption{Case $\delta_1=-0.2$, $\delta_2=0.4$ and
$\tan\beta=8$.}
\end{figure}
\noindent
\begin{figure}[htb]
\epsfxsize=6truecm
\centerline{\epsffile{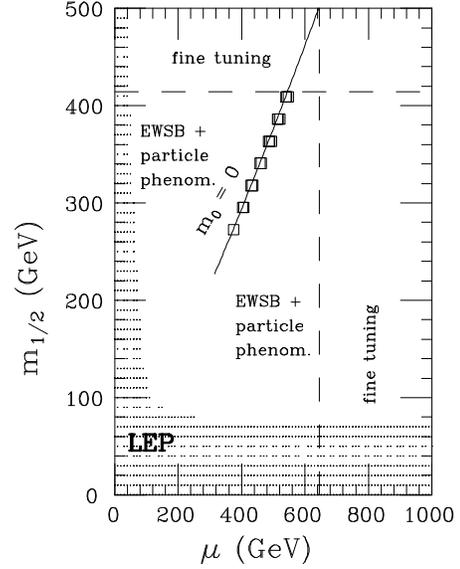}}
\caption{Mass--unification case and $\tan\beta=53$}
\end{figure}
\begin{figure}[htb]
\epsfxsize=6truecm
\centerline{\epsffile{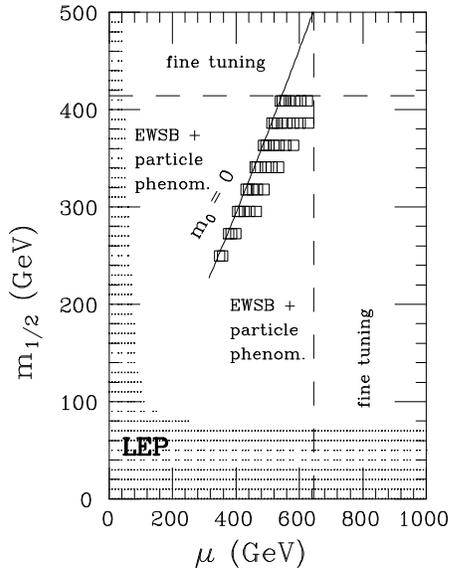}}
\caption{Case $\delta_1=0$, $\delta_2=-0.2$ and 
$\tan\beta=53$.}
\end{figure}
\begin{figure}[htb]
\epsfxsize=6truecm
\centerline{\epsffile{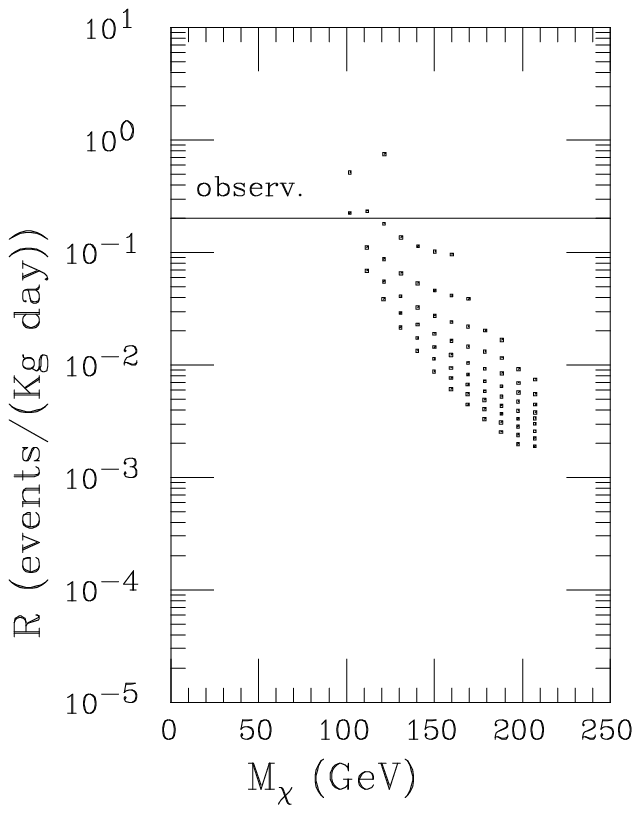}}
\caption{Rate for direct detection ($\delta_1=0,
\delta_2=-0.2$ and $\tan\beta=53$).}
\end{figure}

 This non-universality affects rather strongly the properties of neutralino 
as DM particle: the allowed parameter space regions become larger and 
neutralino is allowed to be Higgsino-dominated, which is favorable for 
detection.

Some results obtained in ref.\cite{Ber95} are illustrated by 
Figs.~2 - 6. The regions allowed for the neutralino as CDM particle are 
shown everywhere by small boxes.

In Fig.2 the regions excluded by the LEP and 
CDF data are shown by dots and labelled as LEP. The regions labelled 
"fine tuning" have an accidental compensation stronger than $1\%$
and thus are excluded.  No-fine-tuning
region inside the broken-line box corresponds to a neutralino mass 
$m_{\chi}\leq 200$~GeV. The region "EWSB+particle phenom." is excluded
by the EWSB condition
combined with particle phenomenology (neutralino as LSP, limits on the 
masses of SUSY particles etc). In the region marked by rarefied dotted lines 
neutralinos overclose the Universe ($\Omega_{\chi}h^2>1$). The solid line 
corresponds to  $m_0=0$. The regions allowed for neutralino 
as CDM particle ($0.01<\Omega_{\chi}h^2<1$) are shown by small boxes. As 
one can see in most regions the neutralinos are overproduced. The allowed regions 
correspond to large $\chi\chi$ annihilation cross-section (e.g. due to 
$Z^0$-pole).

\noindent
Fig.~3 and Fig.~2 differ only by universality: in Fig.~3 
$\delta_1=\delta_2=0$ (mass--unification), while in Fig.~4 $\delta_1=-0.2$
and $\delta_2=0.4$. The allowed region in Fig.~3
 becomes much larger and is shifted into the Higgsino dominated region.
Figs.~4 and 5 are given for $tan\beta=53$.  This large value of $\tan\beta$  
correspond to 
$b-\tau-t$ unification of the Yukawa coupling constants. Again one can notice
that in the mass--unification case ($\delta_1=\delta_2=0$) only a 
small area is allowed for neutralino as CDM particle, while in non-universal 
case ($\delta_1=0, \delta_2=-0.2$) the allowed area becomes larger and 
shifts into the gaugino dominated region.

In Fig.~6 the scatter plot for the rate of direct detection with the 
$Ge$ detector \cite{Beck94} is given for the non-universal 
case ($\delta_1=0, \delta_2=-0.2$) and 
$\tan\beta=53$. We notice that, for some configurations, the experimental
sensitivity is already at the level of the predicted rate.

\section{\bf Conclusions}

1. The density of CDM needed for most cosmological models is given by 
$\Omega_{CDM}h^2=0.2 \pm 0.1$.
There are four candidates for CDM, best motivated from point of view of 
elementary particle physics: neutralino, axion, axino and majoron. 

2. There are different approaches to study the {\em neutralino} as DM particle 
in SUSY models with R-parity conservation.

In the {\em phenomenological approach}, apart from the LEP-CDF  
limits, very few other constraints 
are imposed. In the Minimal Supersymmetric Model  only one GUT relation 
between gauginos masses,
$M_1:M_2:M_3=\alpha_1:\alpha_2:\alpha_3$, is used. While 
the coupling constants are known, the mass parameters needed for calculations 
are taken as the free parameters. Many allowed configurations in the parameter 
space give the neutralino as DM particle with observable signals for direct and 
indirect detection (antiprotons and positrons in our Galaxy and high energy 
neutrinos from the Sun and Earth).

The other extreme case, the complete SUSY SU(5) model with fixed particle 
content, with meeting of coupling constants at the GUT point and with 
the constraints due to proton decay, leaves very little space for the 
neutralino as DM particle.

In the third option the SUSY soft breaking terms (\ref{eq:sbt}) and 
induced EW symmetry breaking are used as the general theoretical framework.
Combined with a no-fine-tuning condition this framework  already gives
essential restrictions. In particular, fine tuning allowed at the level larger
than $1\%$ results in the neutralino being lighter than 200~GeV. Here the 
neutralino is 
gaugino dominated, which is unfavorable for direct detection. If we employ 
further  other restrictions, such as exact unification of gauge 
coupling constants and soft--breaking parameters
at the GUT scale, $b \to s\gamma$ limit and $b-\tau$ unification,
the model becomes as rigid as the one considered above. 

If, on the other 
hand, we relax some constraints, {\em e.g.} by assuming non-universality 
of the scalar mass term in Eq.(\ref{eq:sbt}), then even in the case of $1\%$ 
no-fine-tuning condition the neutralino can be the DM particle in a large area 
of the parameter space and can be detected in some parts of this area in 
direct and indirect experiments.

The neutralino is an observable particle. It can be observed directly in the 
underground experiments, or indirectly, mostly due to the products of 
neutralino-neutralino annihilation.

3. In the framework of supersymmetric theory, the PQ-mechanism for solving 
the problem of strong CP violation, results in a Goldstone 
supermultiplet which contains {\em axion} and its fermionic superpartner, 
{\em axino}.
Both of them can be CDM particles. The most important parameter
here is the scale of PQ symmetry breaking $f_{PQ}$ which is observationally 
constrained as $2\cdot 10^9~GeV <f_{PQ}<8\cdot 10^{11}~GeV$. The axion can be 
the CDM particle, if its mass is $10^{-5} - 10^{-3}~eV$ ( the corresponding 
values of $f_{PQ}$ are between  $2\cdot 10^9~GeV$ and $6\cdot10^{10}~GeV$).
For larger values of $f_{PQ}$ the axino can be CDM particle if a 
mechanism of SUSY breaking provides its mass within the interval 
$10 -100~keV$. The axino can provide both CDM and HDM components needed 
to fit the
 cosmological observations.  The axion can be directly observed 
(e.g. in microwave cavity
experiments) while the axino dark matter is practically unobservable.

4. The Majoron with keV mass can be the warm DM particle which explains 
the galactic scale ($M\sim 10^{12}M_{\odot}$) in the structure formation  
problem. The decay of the Majoron to two photons can produce an observable 
X-ray line in the cosmic background radiation.

{\bf Acknowledgements} \hfill \break
The main results presented here on the neutralino are based on a work
carried out with John Ellis, Nicolao Fornengo and Stefano Scopel. We wish to 
express our thanks to them for collaboration and discussions.
Partial financial
support was provided by the Theoretical Astroparticle Network under contract
No. CHRX--CT93--0120 of the Direction General of the EEC.


\begin{thebibliography}{99}
\bibitem{St94} N. Hata, R. Schrerer, G. Steigman, D. Thomas and T.P. Walker,
preprint OSU-TA-26/94 (astro-ph/9412087),1994
\bibitem{Kly} A. Klypin, S. Borgani, J. Holtzman and J. Primack,
{\em Ap.J.} {\bf444} (1995) 1.
\bibitem{Pri95} J.R. Primack, to be published in Proc. {\em Snowmass} 
(eds E.W. Kolb and R. Peccei) (1995).
\bibitem{PQ77} R.D. Peccei and H.R. Quinn, {\em Phys. Rev. Lett.} {\bf 38} 
(1977) 1440.
\bibitem{KT90} E.W. Kolb and M.S. Turner, {\em The Early Universe}, Addison 
Wesley Company, 1990.
\bibitem{Raf95} G. Raffelt, preprint hep-ph/9502358, 1995. 
\bibitem{Bat94} B.A. Battye and E.P.S. Shellard, {\em Nucl. Phys.} {\bf 423}
(1994) 260.
\bibitem{BaSh94} B.A. Battye and E.P.S. Shellard, preprint DAMTP--R--94--31
(1994) and astro-ph/9408035.
\bibitem{axdet} C. Hagmann et al., {\em 30$^{th}$ Rencontres de Moriond} (1995)
and astro--ph/9508013; \hfill \break
K. Van Bibber et al., preprint UCRL--JC--118357 (1994);\hfill \break
L. Cooper and G.E. Stedman, {\em Phys. Lett.} {\bf B357} (1995) 464.    
\bibitem{Dav86} R. Davis, {\em Phys.Lett.}{\bf B 180}(1986)225
\bibitem{TW82} K. Tamvakis and D. Weiler,{\em Phys. Lett.} {\bf B112} (1982) 
451.
\bibitem{Nie86} J.F. Nieves, {Phys. Rev.} {\bf D33} (1986) 1762.
\bibitem{ChL95} E.J. Chun and A. Lukas, preprint hep-ph/9503233 (1995).
\bibitem{GY92}  T. Goto and M. Yamaguchi, {\em Phys. Lett.} {\bf B276}
(1992) 123.
\bibitem{BGM89} S.A. Bonometto, F. Gabbiani and A. Masiero, {\em Phys.Lett.}
{\bf B139} (1989) 433.
\bibitem{RTW91} K. Rajagopal, M.S. Turner and F. Wilczek, {\em Nucl. Phys.}
{\bf B358} (1991) 447.
\bibitem{BGM94} S.A. Bonometto, F. Gabbiani and A. Masiero, {\em Phys.Rev.}
{\bf D49} (1994) 3918.
\bibitem{Akh92} E. Akhmedov, Z.G. Berezhiani, R.N. Mohapatra and G. Senjanovic,
{\em Phys. Lett.} {\bf B299} (1993) 90. 
\bibitem{sec92} J.Z. Rothstein, K. Babu and D. Seckel, {\em Nucl. Phys.} {\bf
B403} (1993) 725. 
\bibitem{CKO93} J.M. Cline, K. Kainulainen and K. Olive, {\em Astroparticle
Physics} {\bf 1} (1993) 387.
\bibitem{BV93} V. Berezinsky and J.W.F. Valle, {\em Phys. Lett.} {\bf B318}
(1993) 360.
\bibitem{DPV95} A.D. Dolgov, S. Pastor and J.W.F. Valle, preprint FTUV--95--14
(1995) and astro--ph/9506011. 
\bibitem{Bott94} A. Bottino, V. de Alfaro, N. Fornengo, G. Mignola and 
S. Scopel,
{\em Astroparticle Physics} {\bf 2} (1994) 77.
\bibitem{Bott95} A. Bottino, C. Favero, N. Fornengo and G. Mignola,
{\em Astroparticle Physics} {\bf 3} (1995) 77.
\bibitem{JKG} G. Jungman, M. Kamionkowski and K. Griest, to be published in
{\em Phys. Rep.}(1995).
\bibitem{Tur} K. Griest, M. Kamionkowski and M.S. Turner,{\em Phys.Rev.} 
{\bf D41} (1990) 3565;\\ 
M. Kamionkowski and M.S. Turner, {\em Phys. Rev.} {\bf D43} (1991) 1774.
\bibitem{Ka91} M. Kamionkowski, {\em Phys. Rev.} {\bf D44} (1991) 3021.
\bibitem{Bott95a} A. Bottino, N. Fornengo, G. Mignola and L. Moscoso,
{\em Astroparticle Physics} {\bf 3} (1995) (65).
\bibitem{Nan92} J. Lopez, D. Nanopoulos and A. Zichichi, {\em Phys. Lett.}
{\bf B291} (1992) 255.
\bibitem{Nan93} J. Lopez, D. Nanopoulos and H. Pois,{\em Phys. Rev.}{\bf D47}
(1993)2468.
\bibitem{Kane94} G.L. Kane, C. Kolda, L. Roszkowski and J.D. Wells,
{\em Phys. Rev.} {\bf D49} (1994) 6173.
\bibitem{Diel} E. Diehl, G.L. Kane, C. Kolda and J.D. Wells, Michigan Univ.
Preprint, hep-ph/9502399 (1995).
\bibitem{NA93} P. Nath and R. Arnowitt, {\em Phys. Rev. Lett.} {\bf70} (1993) 
3696.
\bibitem{AN95} R. Arnowitt and P. Nath, EPS conference (1995) to be published. 
\bibitem{RS95} R. Rattazzi and U. Sarid, Stanford Univ. preprint (1995).
\bibitem{Pok95} M.Olechowski and S.Pokorski, {\em Phys. Lett.} {\bf B344}
(1995) 201.
\bibitem{Ber95} V. Berezinsky, A. Bottino, J. Ellis, N. Fornengo, G. Mignola
and S. Scopel, {\em Astroparticle Physics} to be published (1995).
\bibitem{IR82} L.E. Ibanez and G.G. Ross, {\em Phys. Lett.}  {\bf B110} 
(1982) 215;\\
J. Ellis, D. Nanoupos and K. Tamvakis, {\em Phys. Lett.} {\bf B121} (1983) 123.
\bibitem{Beck94} M.Beck (Heidelberg-Moscow collaboration) {\em Nucl.
Phys.} {\bf B} (Proc.Suppl.){\bf35} (1994) 150.
\end{thebibliography}
\end{document}